\documentclass{aa}  

\usepackage{graphicx}
\usepackage{txfonts}

\usepackage{caption}
\usepackage{subcaption}
\usepackage{xcolor}

\usepackage{soul}
\setstcolor{red} 

\newcommand{\gaia}{{\it Gaia}\xspace}
\newcommand{\hst}{{\it HST}\xspace}
\newcommand{\gh}{{\it GH}\xspace}

\usepackage{hyperref}
\hypersetup{breaklinks=true,
            colorlinks,
            filecolor=blue,
            linkcolor=blue,
            citecolor=blue,
            urlcolor=blue,
            anchorcolor=blue} 
\begin{document}

   \title{The {\it Hubble} Missing Globular Cluster Survey }

   \subtitle{II. Survey membership tools and kinematic analysis of NGC 6749}
   
   \titlerunning{{\it Hubble} MGCS. II}

   \author{L. Rosignoli\inst{1, 2}\and
          M. Libralato\inst{3}
          \and
          R. Pascale\inst{1}
          \and
          D. Massari\inst{1}
          \and
          E. Dalessandro\inst{1}
          \and
          E. Ceccarelli\inst{1, 2}
          \and
          H. Baumgardt\inst{4}
          \and
          M. Bellazzini\inst{1}
          \and  
          A. Bellini\inst{5}    
          \and
          F. Aguado-Agelet\inst{6,7}
          \and
          S. Cassisi\inst{8, 9}
          \and
          M. Monelli\inst{8, 10}
          \and
          A. Mucciarelli\inst{2, 1}
          \and
          E. Pancino\inst{11}
          \and
          M. Salaris\inst{12, 8}
          \and 
          E. Dodd\inst{13}
          \and 
          F. R. Ferraro\inst{2}
          \and
          B. Lanzoni\inst{2}
          }

   \institute{INAF - Astrophysics and Space Science Observatory of 
              Bologna, Via Gobetti 93/3, 40129 Bologna, Italy\\ \email{luca.rosignoli@inaf.it}
         \and
             Department of Physics and Astronomy, University of 
             Bologna, Via Gobetti 93/2, 40129 Bologna, Italy 
         \and
            INAF, Osservatorio Astronomico di Padova, Vicolo dell’Osservatorio 5, Padova, I-35122, Italy
        \and
            School of Mathematics and Physics, University of Queensland, Brisbane, Qld 4072, Australia
         \and 
            Space Telescope Science Institute, 3700 San Martin Drive, Baltimore, MD 21218, USA
         \and
             atlanTTic, Universidade de Vigo, Escola de Enxeñar\'ia de Telecomunicaci\'on, 36310, Vigo, Spain 
         \and
             Universidad de La Laguna, Avda. Astrof\'isico Fco. S\'anchez, E-38205 La Laguna, Tenerife, Spain
         \and
             INAF – Osservatorio Astronomico di Abruzzo, Via M. Maggini, 64100 Teramo, Italy
         \and
             INFN - Sezione di Pisa, Universit\'a di Pisa, Largo Pontecorvo 3, 56127 Pisa, Italy
         \and 
            INAF – Osservatorio Astronomico di Roma, Via Frascati 33, 00078 Monte Porzio Catone, Roma, Italy
         \and
            INAF - Osservatorio Astrofisico di Arcetri, Largo E. Fermi 5, I-50125 Firenze, Italy
         \and
            Astrophysics Research Institute, Liverpool John Moores University, 146 Brownlow Hill, Liverpool L3 5RF, UK     
         \and
             Institute for Computational Cosmology \& Centre for Extragalactic Astronomy, Department of Physics, Durham University, South Road,
             Durham, DH1 3LE, UK
        }
   \date{today}

 
  \abstract
   {The {\it Hubble} Missing Globular Cluster Survey has secured high-quality astro-photometric data in two bands for 34 clusters never observed with {\it HST}. When combined with {\it Gaia} positional measurements, this data set enables the investigation of the bulk motion and the internal kinematics of these poorly studied clusters to an unprecedented level of detail.
   Focusing on the case of NGC~6749, we here showcase how the combined {\it Gaia-HST} proper motions have a quality sufficient to accurately assess the cluster stellar membership, determine its absolute proper motion with a precision superior to {\it Gaia}, and to investigate its kinematic profile for the first time. 
   Proper motions are determined using the public code \texttt{GAIAHUB}, which for NGC~6749 combines data sets separated in time by $\sim8$ years.
   The resulting measurements improve the precision of {\it Gaia} proper motions by a factor of 10 at the faint end, and enable recovering the proper motion for 662 stars for which {\it Gaia} could only measure the positions. These proper motions are efficient in decontaminating the colour-magnitude diagram of NGC~6749, and make it possible to compare the efficacy of a method of statistical decontamination  that relies only on the photometric information extracted from the {\it HST} parallel fields. Finally, using the sample of best-measured proper motions we determine the velocity dispersion and anisotropy profiles of NGC~6749, that reveal an isotropic behaviour in the cluster inner regions and a slight radial anisotropy outside 1.5 half-light radii. 
   The proper motions and the code to statistically decontaminate the cluster's color-magnitude diagram are made available as public products of the survey. }


   \maketitle
%
\section{Introduction}
Globular clusters (GCs) are exceptional laboratories to understand fundamental processes of galaxy formation and evolution \citep{searle78}. The system of Milky Way (MW) GCs is no exception in this sense, as its chemical \citep[see e.g.,][]{recio-blanco18, horta20, monty24, ceccarelli24a}, dynamical \citep[e.g.,][]{massari19, callingham22, chen24} and chronological \citep[e.g.,][]{de_angeli_galactic_2005, marin-franch09, dotter10, leaman13, massari23} properties allow us to investigate the history of our Galaxy in detail.

Historically, the most fundamental advancements in the study of the system of MW GCs came from observational campaigns of the \textit{Hubble Space Telescope} (\textit{HST}). The primary examples of these are the ACS Survey of Galactic Globular Clusters \citep{sarajedini07}, and the HST UV Legacy Survey of Galactic Globular Clusters \citep{piotto15}, which have shaped the state-of-the-art of GC science. The latest example is the {\it Hubble} Missing Globular Cluster Survey (MGCS, HST Treasury Program GO-17435, PI: Massari), a campaign that has just been completed and that aims at providing homogeneous two-bands photometry and astrometry for 34 GCs that have never been observed with {\it HST} before. 
The characteristics of the survey, the targets and first results have been presented in \citet[][Paper I, hereafter]{massari_2025mgcs}. There the authors showed that the high photometric precision achieved by the survey in both optical and near-IR bands enabled the determination of the age of two GCs with sub-Gyr precision, despite one of these (namely 2MASS-GC01) being the highest-extincted GC currently known. Incidentally, the first results of the survey also revealed that 2MASS-GC01 could be the youngest GC known to date, even if its age of $\sim7$ Gyr is so low as to question its nature as a genuine globular.

While the astro-photometric catalogues of all the 34 observed targets will be made publicly available with a forthcoming publication (Libralato et al., in prep.), this paper focuses on one cluster, namely NGC~6749, and describes its analysis by means of two techniques, both of which are crucial to achieve the objectives of the survey.

The first part of the investigation relies on measurements of the proper motions (PMs) for the stars in common between the {\it HST} MGCS and {\it Gaia} DR3 data \citep{gaiadr3}. PM-based kinematic analyses of GCs have flourished in the last decade thanks to {\it HST} and {\it Gaia}. This astrometric Renaissance has greatly improved our understanding of GCs, clearly highlighting their kinematic complexity \citep{bianchini_internal_2018, sollima_eye_2019, vasiliev_2021,libralato_2022,watkins22, haeberle24, dalessandro_3d_2024}. Nevertheless, kinematic information is not always available for all GCs with {\it HST} or {\it Gaia} alone because of technical limitations like crowding, extinction, distance (for {\it Gaia}), or simply the lack of sufficient multi-epoch data (for {\it HST}). The combination of {\it HST} and {\it Gaia}, though, has shown that it is possible to bypass some of these limitations and obtain PMs for the subsample of stars common to both catalogs \citep[see e.g.,][]{massari18, massari20, delpino22, bennet_2024, libralato24}, opening new possibilities to study additional, so-far-poorly-studied, systems.

While the {\it HST}-{\it Gaia} duo provides a powerful new tool, it is not without its own limitations. Some of the goals of the MGCS survey, such as the determination of age, mass function, and binary fraction, require a clean sample of cluster stars spanning a wide range of magnitudes/masses along the main sequence (MS). However, our {\it HST} images can observe objects that are 5--6 magnitudes fainter than the {\it Gaia} limit. Consequently, relying only on {\it HST}-{\it Gaia} PMs would unavoidably lead to the loss of a significant fraction of objects along the faint MS of our target clusters. For this reason, we have devised an additional analysis based on the statistical decontamination of the CMD, taking advantage of the simultaneous \hst observations in the core and in an external field.

In this paper, we showcase the application of these two techniques to the case of NGC~6749, an in-situ disc GC \citep[][eDR3 edition]{massari19} located at a distance of 7.6 kpc \citep[][]{baumgardt21}.  It is a relatively massive GC 
with a half-mass relaxation time of $\sim$2.5 Gyr \citep{baumgardt21}. Given the size of this GC \citep[half-light radius of 66"][2010 edition]{harris96}, we were able to map the kinematics of this cluster out to about two half-light radii with a single \hst field in the core.

In Section \ref{sec:data_reduction} we present the dataset together with photometric calibration (Section \ref{subsec:data_red_photometry}) and the PM estimation (Section \ref{subsec:data_red_astrometry}). Section \ref{sec:membership} describes the two methods to infer membership probabilities of each star, one using  \hst-\gaia PMs (Section \ref{subsec:kinematic_membership}) and another focused on the statistical approach (Section \ref{subsec:photometric_membership}). In Section \ref{sec:kinematic_analysis}, we validate the model used to study the PMs firstly with a mock catalog and then by comparing our results with the literature (Section \ref{subsec:mock_validation}). After the validation, we used the \hst-\gaia PMs to perform the radial kinematic analysis of NGC~6749 (Section \ref{subsec:radial_analysis}), and we derived the physical properties through N-body simulation comparison in Section \ref{subsec:nbody}. 
Finally, we summarize this work in Section \ref{sec:conclusions}.

\section{Data reduction}\label{sec:data_reduction}
\begin{figure}[!t]
    \centering
    \includegraphics[width=\columnwidth]{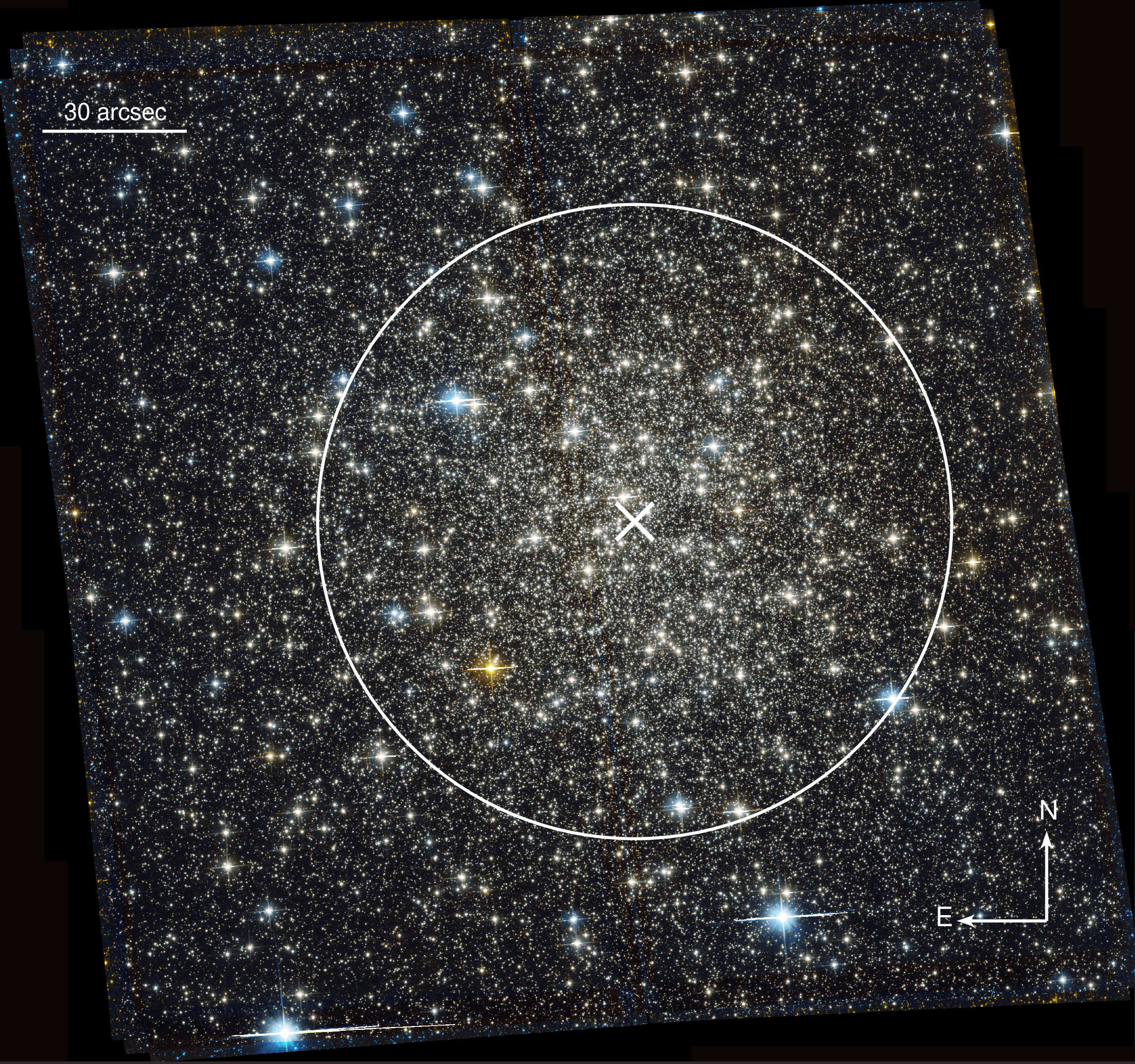}
    \caption{Colored image of NGC~6749 using the new \hst (ACS/WFC) observations. The white cross shows the center of the cluster \citep[according to][]{baumgardt21}, while the white circle has a radius equal to the half-light radius.}
    \label{fig:NGC6749_rgb}
\end{figure}

We made use of the \hst images of NGC~6749 taken within the MGCS on 2024, March 9--12. The data set consits of four exposures (2$\times$699\,s, 1$\times$337\,s, 1$\times$30\,s) per filter collected with the Wide Field Channel (WFC) of the Advanced Camera for Surveys (ACS) in the F606W and F814W bands on target, and of an analog set of parallel exposures with identical setup using the Ultraviolet and Visible Instrument (UVIS) of the Wide Field Camera 3 (WFC3). The ACS/WFC data cover the central region of the cluster as shown in Fig.~\ref{fig:NGC6749_rgb}, whereas the WFC3/UVIS parallel fields are off-center by about 7 arcmin.

    \subsection{Photometry}\label{subsec:data_red_photometry}
     The reduction of this data set was performed as described in \citet{massari_2025mgcs}, in accordance with the well-established prescriptions outlined in a number of papers presenting high-precision astrometry and photometry with {\it HST} images \citep{2017BelliniwCenI,nardiello2018,2018LibralatoNGC362,libralato_2022}. 

    We refer the reader to Paper I, and to the forthcoming paper by Libralato et al., for the details of the astro-photometric analysis. Briefly, the data reduction consists of two steps called \textit{first-} and \textit{second-pass} photometry. The first-pass photometry is used to create the initial astro-photometric catalogs, one for each \textit{HST} image\footnote{We made use of \texttt{\_flc} exposures, i.e., dark- and bias-corrected, flat-fielded, not-resampled images that were pipeline-corrected for the charge-transfer-efficiency defects \citep{2018acs..rept....4A}.}. Positions and magnitudes in these catalogs were obtained via the software \texttt{hst1pass}\footnote{The code and its ancillary files can be found here: \href{https://www.stsci.edu/~jayander/HST1PASS/}{https://www.stsci.edu/$\sim$jayander/HST1PASS/}.} \citep{2022acs..rept....2A,2022wfc..rept....5A} that performs effective-point-spread-function (ePSF) fit by means of the publicly-available \hst ePSF models appropriately fine-tuned for each image. Positions are corrected for the effect of geometric distortion using publicly-available corrections\footnote{Can be found in \href{https://www.stsci.edu/~jayander/HST1PASS/LIB/GDCs/STDGDCs/}{https://www.stsci.edu/$\sim$jayander/../STDGDCs/}}. These first-pass catalogs are used to set up the astrometric and photometric reference systems for the second-pass photometry. The second-pass photometry is obtained with proprietary software \texttt{KS2} \citep[Anderson in prep.;][]{2017BelliniwCenI}, which analyzes all input images at once (thus enhancing the detection of faint sources) and is designed to overcome crowding-related limitations typical of environments like GCs.

    \subsection{Astrometry}\label{subsec:data_red_astrometry}
    The astrometric analysis of NGC~6749 focuses only on the central field observed with the ACS/WFC. 
    The PMs of the stars in this region are computed by means of the publicly-available tool \texttt{GAIAHUB} \citep{del_Pino_2022}. \texttt{GAIAHUB} combines \textit{HST} and \textit{Gaia} data to obtain PMs with a temporal baseline longer than that of the \textit{Gaia}-DR3 PMs. As shown by, e.g., \citet{del_Pino_2022} and \citet{bennet_2024}, \texttt{GAIAHUB} delivers PMs that can be one order of magnitude more precise than \textit{Gaia}'s at the \textit{Gaia} faint-end and in crowded environments.

    \texttt{GAIAHUB} relies on astro-photometric catalogs computed for each \hst image by means of the first-pass-photometry tool \texttt{hst1pass}, which is run in the background at the beginning of the process. To ensure a self-consistent infrastructure throughout our project, we modified \texttt{GAIAHUB} to use our first-pass astro-photometric catalogs (Sect.~\ref{subsec:data_red_photometry}) instead of producing its own.
    
    Before computing PMs, \texttt{GAIAHUB} transforms \hst positions onto the reference frame of \gaia via a six-parameter linear transformation (two offsets, one rotation, one change of scale, and two skew terms\footnote{Skew terms represent the deviation from orthogonality and the relative scale difference between the two axes.}). In general, the coefficients of such transformations are computed by comparing two sets of positions of the same reference stars. If these positions are taken at different epochs, these transformations would map both the different alignment between the frames and the actual motion of the reference stars between the two epochs. In our case, the latter contribution would bias our frame alignment because cluster and field stars have significantly different motions. To solve for this problem, one could either select a single group of stars (e.g., cluster stars, whose velocity dispersion and coherent motion would imprint a minimal contribution in the transformations), or propagate the positions of one of the two catalogs at the epoch of the other catalog (so that all stars can be used, regardless the population they belong to). We choose the latter option (using the \texttt{\mbox{-{}-rewind\_stars}} flag) so to avoid a membership selection that at this stage could bias our PMs\footnote{For completeness, we also tested the membership option (\texttt{\mbox{-{}-members}} flag) and found a slightly worse result due to the significant overlap between the contaminating field component and the cluster stars that makes the membership selection challenging.}, and to increase statistics.

    The absolute PMs from \texttt{GAIAHUB} of the stars in the central field of NGC~6749 present a series of systematics as a function of position, magnitude and color. We corrected for these systematic errors by comparing our {\it Gaia}-\hst (\gh~ hereafter) PMs with those in the \gaia DR3 catalog. A detailed description of the systematic corrections is provided in Appendix~\ref{appendix:pms_calibration}.
    
\begin{figure*}[!htb]
        \centering
        \includegraphics[width=\textwidth]{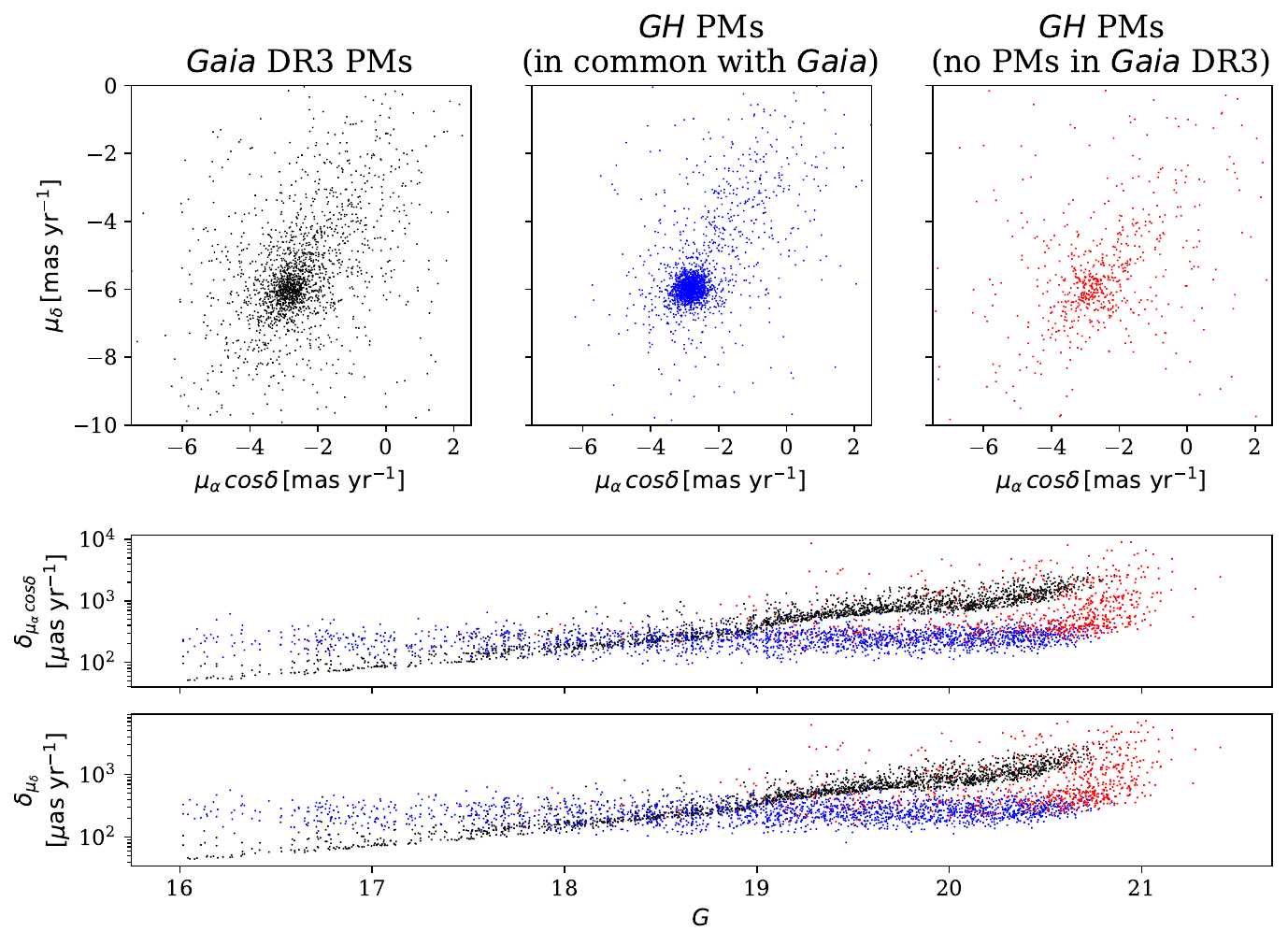}
        \caption{Comparison between the \gaia and \gh PMs. The top panels show the VPDs of the PMs while the two bottom panels show the PM errors along $\alpha\cos\delta$ and $\delta$, as a function of Gaia \textit{G} magnitude. The black dots refer to the \gaia PMs, while blue and red dots show stars in our catalog with and without a corresponding PM in the \gaia catalog, respectively.}
        \label{fig:pms_comparison}
    \end{figure*}

    A comparison between our {\it GH} and \textit{Gaia}'s PMs is shown in Fig.~\ref{fig:pms_comparison}. The top panels present the vector-point diagrams (VPDs) of the PMs in the \gaia DR3 catalog (left panel) and our PMs (middle and right panels). While the middle panel presents objects with a counterpart entry in the \gaia catalog, the right panel includes only sources without a PM in the \gaia catalog. The comparison of the VPDs for the same stars in our and \gaia panels (middle and left panels, respectively) already provides a qualitative assessment of the improvement brought by the \textit{HST}-\gaia synergy, as the more populated clump in these diagrams (i.e., the distribution of the cluster stars) looks tighter with our PMs than with those in the \gaia DR3 catalog. This improvement in precision is more quantitatively shown by the two bottom panels in Fig.~\ref{fig:pms_comparison} where we compare the PM errors in each component of the motion as a function of \textit{G} magnitude. At magnitudes fainter than $G \simeq 18.5$, our PMs become more precise than the ones from \gaia by a factor of 3 and up to 10 at the very faint end, with a median value of $\delta_{\mu} = 0.42\,\,\mathrm{mas}/\mathrm{yr}$ for \gh and $\delta_{\mu} = 1.09\,\,\mathrm{mas}/\mathrm{yr}$ for \gaia\footnote{For magnitude below $G\simeq18.5$ the ``raw'' \gh PMs are still better than the ones of \gaia. In particular, the median error value for \gh is $\delta_{\mu} = 0.04\,\,\mathrm{mas}/\mathrm{yr}$ compared to $\delta_{\mu} = 0.22\,\,\mathrm{mas}/\mathrm{yr}$ for \gaia, but after the calibration described in Section \ref{appendix:pms_calibration} the error increase due to the square-root summation of the initial error with the uncertainty on the correction.}. Furthermore, the \texttt{GAIAHUB} catalog contains PMs for 2634 stars compared to the 1972 from \gaia, incrementing by 662 the number stars with this information. This \textit{de-facto} extends the astrometry to stars $\sim$0.5 mag fainter than in the \gaia catalog.

\begin{figure*}[htb]
       \centering
        \includegraphics[width=\textwidth]{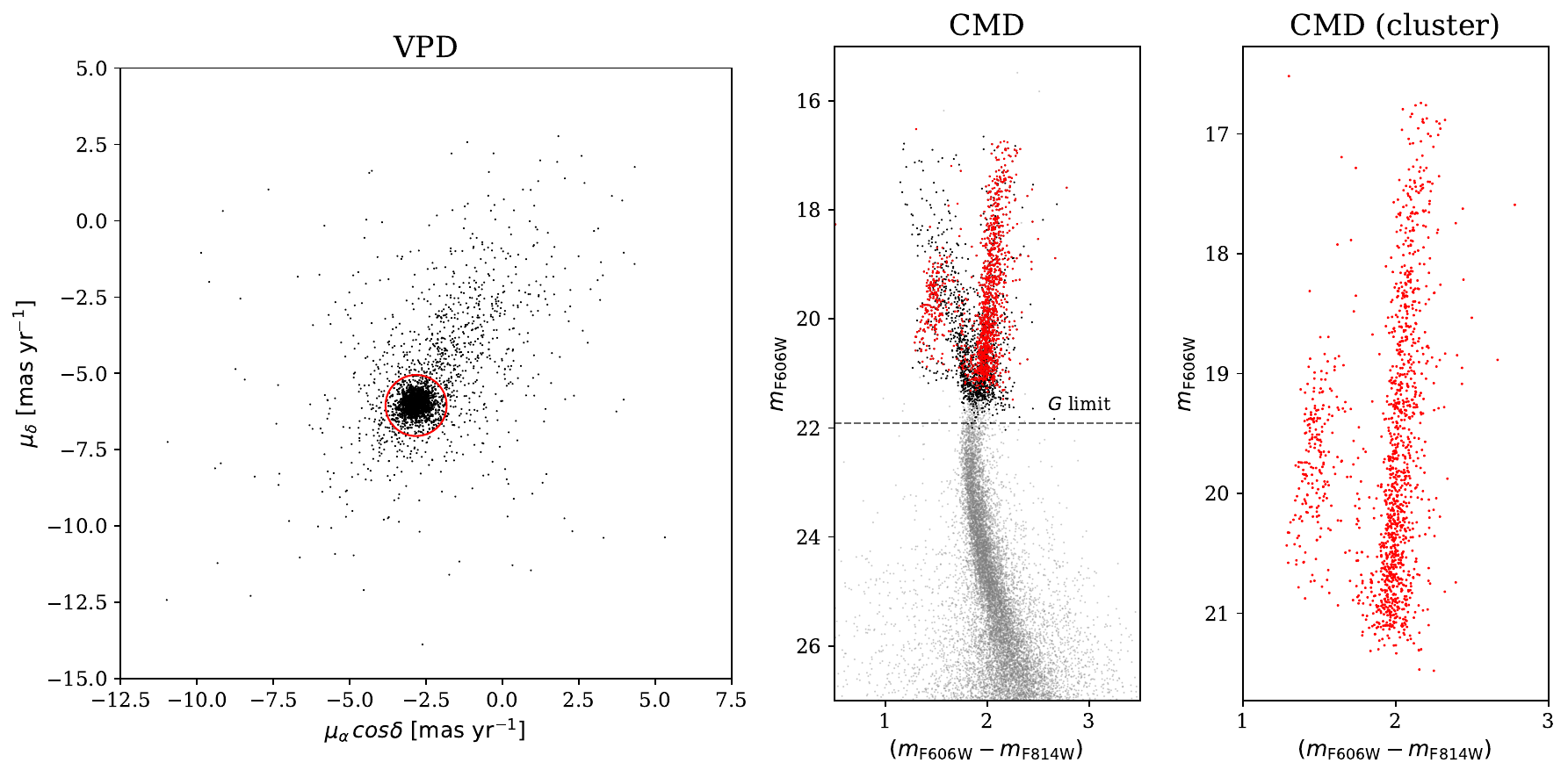}
        \caption{VPD (left panel) and CMDs (middle and right panels) of NGC~6749 using the new \hst data of the MGCS survey. The red circle with radius 1 $\mathrm{mas}\,\mathrm{yr}^{-1}$ in the VPD is used to select likely cluster stars  (those included within the circle). In the central CMD, the gray dots represents all stars in our \hst photometric catalog, while black and red dots are, respectively, foreground and cluster stars for which we have a PM measurements. The horizontal dashed line corresponds to the \gaia limiting magnitude. Finally, the right panel shows the location of cluster member stars in the CMD.}
        \label{fig:vpd_and_cmd_6749}
    \end{figure*}

     To reject bad measurements\footnote{Here we focused only on well-measured sources identified photometrically by means of a mix of percentile-based selections on the quality of the ePSF fit, magnitude rms and ``stellarity'' index, number of images used in the flux measurement, impact of the neighbour flux and local sky \citep[see, e.g.,][and Libralato et al., in prep]{libralato_2022, massari_2025mgcs}.}, we applied the following criteria:
    \citep{bellini_2014,libralato_2018,libralato_2022}:
    
    \begin{equation}
        \delta_{\mu} \leq 0.5 * \sigma_{\mu}^{\rm local}(n)
    \end{equation}
    where $\delta_{\mu}$ is the PM error, and $\sigma_\mu^{\rm local}$ is the local velocity dispersion measured using the $n=20$ closest stars to each source (target excluded).
    Figure~\ref{fig:vpd_and_cmd_6749} shows the vector point diagram (VPD) and the color-magnitude diagram (CMD) of NGC~6749.
    In the VPD, a tight clump of stars located at about ($\mu_{\alpha}cos\delta$, $\mu_{\delta})=(-3,-5.5)$ mas~yr$^{-1}$ is embedded in other sources with a broader distribution. We selected stars within a radius of 1.0 $\mathrm{mas}\,\mathrm{yr}^{-1}$ from the center of the tight clump (red circle) and plotted them in the CMDs (red dots in the middle panel). As expected, these are mainly cluster members (well aligned along the evolutionary sequences in the CMD), whereas the broadly-distributed population in the VPD (those beyond the red circle in the left panel) comprises field stars along the line of sight of the cluster (black dots in the CMD). Zooming in around the CMD location of these stars (right panel), we can identify the cluster's red-giant branch (RGB) and horizontal branche (HB). In the following Section, we address the problem of the cluster membership in a quantitative way, using two different approaches.
    
\section{Membership probability}\label{sec:membership}
In this work, we adopt two independent approaches to estimate the membership probability of the cluster sources detected in the MGCS images. The first approach exploits their PM measurements, while the second relies on the CMD comparison between the target and the WFC3/UVIS parallel observations. In Sections~\ref{subsec:kinematic_membership} and ~\ref{subsec:photometric_membership}  we thoroughly describe each method, while in Section \ref{subsec:membership_comparison} we make a comparison between the two membership selections.
    
    \subsection{Kinematic membership}\label{subsec:kinematic_membership}
    Often, the PM distributions of a GC and the background field overlap significantly \citep[][]{bellini_2014, libralato_2022}. This overlap complicates a clear-cut separation between members and non-members, thereby necessitating a probabilistic approach.
    This approach assumes that the observed sample of stellar PMs is a realization of an underlying model. In its general form, such model consist of $n_{\rm comp}$ components, each corresponding to a distinct stellar population with its own kinematic structure and contributing to the global PMs probability distribution $\mathcal{P}(\vec{\mu})$ with a different weight
    \begin{equation}\label{for:prob}
        \mathcal{P}(\vec{\mu}) = \sum_{i=1}^{n_{\rm comp}}w_i\mathcal{P}_i(\vec{\mu}),
    \end{equation}
    where $\vec{\mu}\equiv(\mu_\alpha\cos\delta,\mu_\delta$), $\mathcal{P}_i(\vec{\mu})$  is the probability distribution of each component, and $w_i$ represents its fractional contribution. We further require that
    \begin{equation}\label{for:wi}
        \sum_{i=0}^{n_{\rm com}} w_i = 1,
    \end{equation}
    to ensure proper normalization. The model parameters are determined within a fully Bayesian framework, by maximizing a well motivated likelihood function of the data. Once the parameters are determined, the model provides a probabilistic framework for assigning to each star a membership probability with respect to the different components.

    In our specific application, we distinguish between two Gaussian components (i.e. a Gaussian Mixture Model, GMM): the target GC and a term describing the foreground contamination (FG). Therefore, the model's probability distribution (\ref{for:prob}) reduces to
    \begin{equation}
        \mathcal{P}(\vec{\mu}) = w_1 G_{\rm GC}(\vec{\mu}) + (1-w_1)G_{\rm FG}(\vec{\mu}),
    \end{equation}
    where $G_{\rm GC}$ and $G_{\rm FG}$ describe the PM distributions of the GC and the foreground stars, respectively, $w_1$ sets the relative contribution of GC over foreground.

    We model each component as bi-variate Gaussian:
    \begin{equation}
    G_i(\vec{\mu}) = \frac{\exp\big[-\frac{1}{2}(\vec{\mu}-\vec{\hat{\mu}}_i)^T\Sigma^{-1}_i(\vec{\mu}-\vec{\hat{\mu}}_i)\big]}{\sqrt{(2\pi)^2\det\Sigma_i}}
    \end{equation}
    where $\vec{\hat{\mu}}_i = (\hat{\mu}_{\alpha,i},\,\hat{\mu}_{\delta,i})$ and $\Sigma_i$ are the center and the covariance matrix of the $i$-th component respectively, with $i$=GC or FG. The covariance matrix $\Sigma_i$, is:
    \begin{equation}
    \Sigma_i=\begin{pmatrix}
            \sigma_{\alpha,i}^2 & \rho_i\sigma_{\alpha,i}\sigma_{\delta,i}\\
            \rho_i\sigma_{\alpha,i}\sigma_{\delta,i}&\sigma_{\delta,i}^2.
        \end{pmatrix}
    \end{equation}
    The model consists of the following free parameters, one per component ($i=$GC or FG):
    \begin{itemize}
        \item $\vec{\hat{\mu}}_i$: the center of motion of each component.
        \item $\vec{\sigma}_i$: the standard deviation of each component.
        \item $\rho_i$: the correlation between the standard deviation components.
        \item $w_1$: the weight of the GC gaussian in the mixture.
    \end{itemize}
    for a total of 6 free parameters.

    The likelihood of the model, given the observed dataset is:
    \begin{equation}
        \mathcal{L} = \prod_j (\mathcal{P}\ast \mathcal{E})(\vec{\mu}_j),
    \end{equation}
    where the product extends over the stars in our sample, $\ast$ represents the convolution with the error function $\mathcal{E}$, which is a bivariate gaussian distribution with null mean, and covariance given by
    \begin{equation}
    \Sigma=\begin{pmatrix}
            \varepsilon^2 & 0\\
            0&\varepsilon^2
        \end{pmatrix}
    \end{equation}

     One of the great advantages of using gaussians when describing multiple components in a mixture model, is that the convolution between each model's component and the error function is still a gaussian with mean and covariance matrix, $\Sigma_i^{obs}$, given by the sums of the individual means and covariance matrices. In our specific case, the resulting covariance matrix can be written as
    \begin{equation}
    \Sigma_i^{obs}=\begin{pmatrix}
            (\sigma_{\alpha,i}^{obs})^2 & \rho_i^{obs}\sigma_{\alpha,i}^{obs}\sigma_{\delta,i}^{obs}\\
            \rho_i^{obs}\sigma_{\alpha,i}^{obs}\sigma_{\delta,i}^{obs}&(\sigma_{\delta,i}^{obs})^2
        \end{pmatrix}
    \end{equation}
    where we have called    \begin{equation}\label{eq:sigma_obs}
        \vec{\sigma}_i^{obs}=\sqrt{\vec{\sigma}_i^2 + \varepsilon^2}
    \end{equation}
    with $\vec{\sigma}_i^{obs}=(\sigma_{\alpha,i}^{obs},\,\sigma_{\delta,i}^{obs})$ and 
    \begin{equation}
        \rho_i^{obs} = \rho_i\frac{\sigma_{\alpha,i}\sigma_{\delta,i}}{\sigma_{\alpha,i}^{obs}\sigma_{\delta,i}^{obs}}.
    \end{equation}
    In the above quations $i=$ GC or FG.

    We determine the models free parameters using a Monte Carlo Markov Chain (MCMC) method, relying on the \texttt{EMCEE} package \citep{Foreman-Mackey_2013}. 
    To sample the posterior and determine the model parameters confidence intervals, we run a MCMC with 50 walkers, each evolved for 10000 steps. We discarded the first 500 steps of each walker as burn-in and applied a thinning factor of 20, consistent with median value of the autocorrelation lengths of the free parameters, thereby ensuring a sample of effectively independent draws. The remaining steps are used to compute the distributions for the model's free parameters. Median values of parameters and any other models' derived quantity are estimated as the 50th percentile of the corresponding distributions; the $1\sigma$ credible intervals are determined using the 16th and 84th percentiles of the distributions.
    
\begin{figure*}
        \centering
        \includegraphics[width=\textwidth]{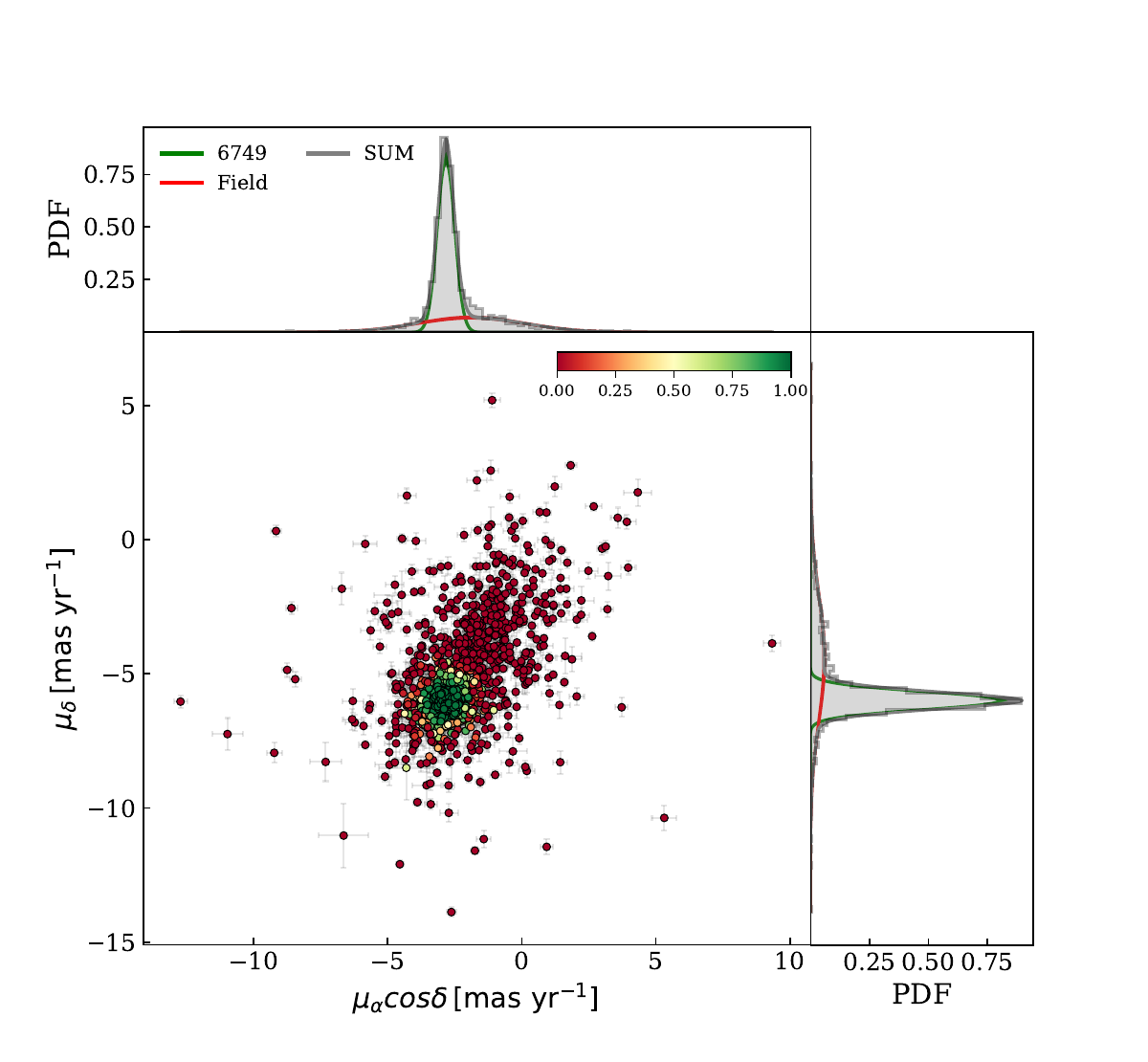}
        \caption{Distribution of the NGC~6749 \gh PMs. The central scatter plot shows the 2D VPD, where the points are color-coded by the membership probability obtained from the MCMC chains (color bar). The top and right histograms show the normalized marginal distributions of the PMs, together with the posterior distributions of the GMM components, the target in red, the field in blue and their sum in gray.}
        \label{fig:vpd_with_membership}
    \end{figure*}

    We assign membership probability to each star in the sample in the following way. For each star $k$, we compute the distribution of $p_k$, i.e. the probability of the $k$-th star to belong the GC. This probability is defined as
    \begin{equation}\label{eq:pm_membership}
        p_k = \frac{ w_1 (G_{\rm GC}\ast\mathcal{E})(\vec{\mu}_k)}{(\mathcal{P}\ast \mathcal{E})(\vec{\mu}_k)}
    \end{equation}
    Consistently with credible interval estimates, we take as probability the 50th percentile of the distribution. 
    Fig.~\ref{fig:vpd_with_membership} shows the VPD of NGC~6749 using the new \gh PMs. The marginal distributions along the two PM axes are displayed as gray histograms together with the individual components (red for the cluster and blue for the field) and their sum (gray). In addition, the points are color-coded based on their membership probability.
    
    \subsection{Photometric membership}\label{subsec:photometric_membership}
\begin{figure*}[tb]
            \centering
            \includegraphics[width=\textwidth]{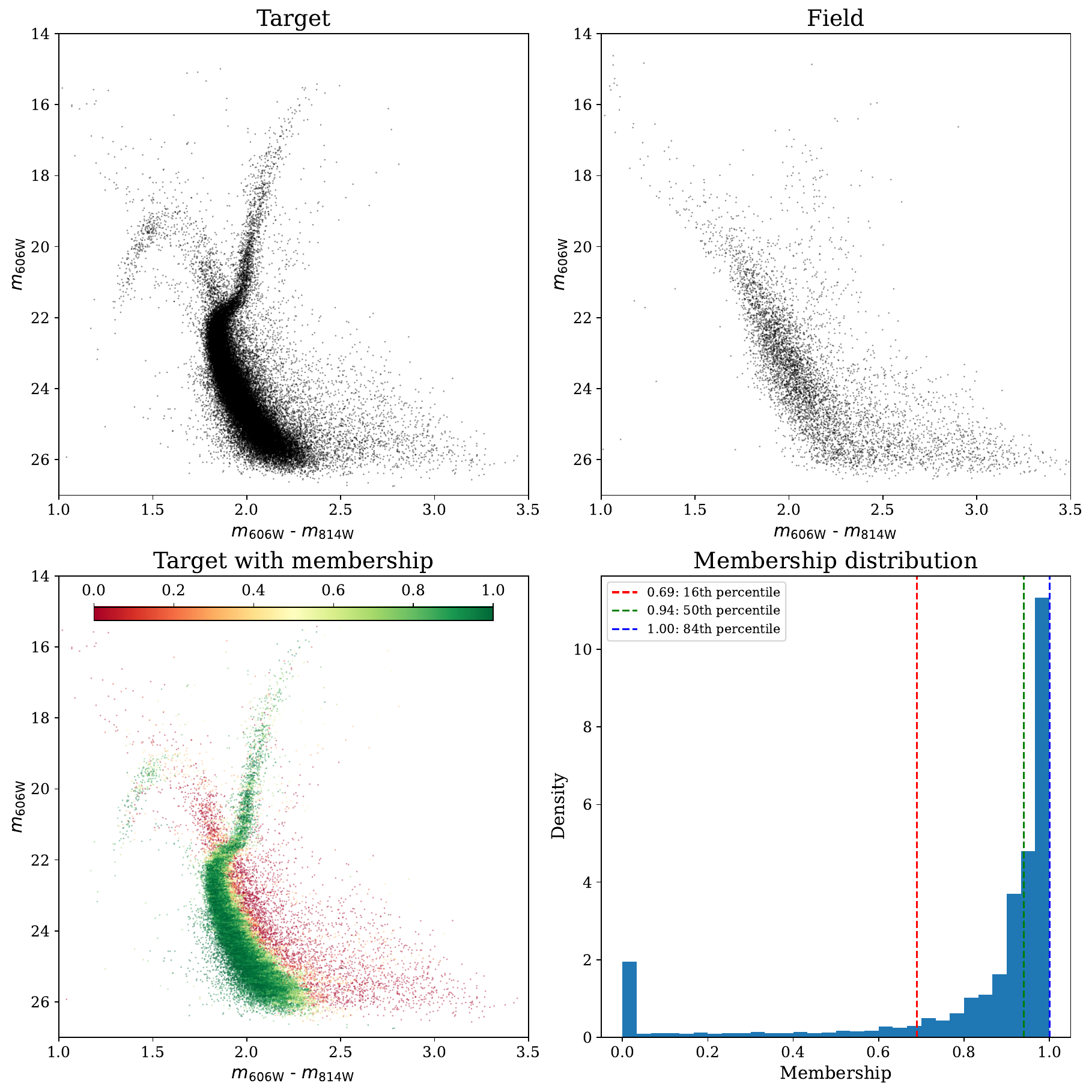}
            \caption{Result of the statistical decontamination. ({\it top-left panel}) Contaminated target CMD, ({\it top-right panel}) parallel field CMD, ({\it bottom-left panel}) target CMD color-coded for the membership probability, ({\it bottom-right panel}) membership probability distribution. The target CMDs shown are corrected for the differential reddening.}
            \label{fig:membership_overview}
    \end{figure*}

    Even though the membership obtained from the PMs is more accurate, the kinematic information is available only for a limited number of stars.
    Another approach exploits the color-magnitude diagram (CMD) to statistically infer the membership. By using the photometric information, the number of suitable stars drastically increase. This method requires the observation of a parallel field, close enough to the target to be representative of the contaminating stellar population, but sufficiently away to exclude GC members. 
    Many algorithms have been proposed over the last 20 years to determine the membership in this way \citep[e.g.,][]{mighell96, knoetig14, cabreraziri16}. The general idea behind these algorithms is that the CMD of the field can be statistically subtracted from the CMD of the target GC. Some approaches define a specific metric to evaluate the space nearby each star, in order to understand its likelihood to be a contaminating star \citep[e.g.][]{milone_2018a}. Other algorithms implement a statistical subtraction to retrieve the membership probability. Usually, these approaches divide the CMDs (both the target and the parallel field) space in cells. For each cell, the information coming from the field CMD is used to perform a statistical subtraction in the same cell of the target CMD. Some implementations use a cell with fixed size \citep{bonatto_2007, dalessandro_double_2019}, while more recent works use adaptive grids.
    Our approach belongs to the latter case, and works as follows. We try to preserve the spatial resolution of the target CMD as much as possible, while keeping a sufficient statistics per cell, to avoid shot noise bias. The grid is created via a Voronoi tesselation that naturally follows the CMD's star density. Moreover, to ensure a uniform spatial distribution of contaminating stars (as we expect from a field population) we split the statistical subtraction in radial bins containing always the same amount of contaminants \citep[see for example][]{dalessandro_double_2019}.
    At the beginning of the process the membership is set to 1.0 for every star, then the following steps are done to update the membership: 
    \begin{enumerate}
        \item Creation of the Voronoi grid, based on the target CMD of the stars contained in the radial bin.
        \item Dilation of the Voronoi grid. By design the Voronoi grid creates a cell for each star. If we perform the following steps with only one star per cell, the final membership distribution will be biased by the shot noise. For this purpose, the Voronoi tesselation is dilated by merging nearby cells, starting from the smallest ones. This process is repeated until the distribution of the number of stars per-cell has a median value equal or greater than a desired threshold. 
        We set a threshold of 10 stars as the best trade-off between the loss of the grid resolution and a good statistic per cell\footnote{The value of 10 stars worked for this cluster, but this number has to be evaluated for each case, depending on the amount of stars available.}.
        \item Count of the number of stars in the field CMD that fall in each cell, $n_{field}$. 
        \item Rescale $n_{field}$ for the ratio between the Field of View (FoV) of the parallel field ($fov_{field}$) and the sky area covered by the current radial bin ($fov_{bin}$), $n_{sub}=n_{field} \cdot \frac{fov_{bin}}{fov_{field}}$ 
        \item Convert $n_{sub}$ from a decimal to integer number with the following criteria: 
        \begin{equation*}
        n_{sub}' =
        \begin{cases}
        \mathcal{B}([1 - n_{sub},\ n_{sub}]), & \text{if } n_{sub} < 1 \\
        \text{int}(n_{sub}) + \mathcal{B}([1 - \text{dig}(n_{sub}),\,\text{dig}(n_{sub})]), & \text{if } n_{sub} \geq 1
        \end{cases}
        \end{equation*}
        where $\text{int}(n_{sub})$ and $\text{dig}(n_{sub})$ are the mantissa and the decimal digits respectively and $\mathcal{B}$ is the Bernoulli distribution. In this way we try to take into account the decimal part of the normalized $n_{sub}$.
        \item For each cell, $n_{sub}'$ target stars are randomly subtracted. If $n_{sub}'$ is greater than the number of target stars within the cell then all the stars are subtracted. This step is repeated a large number of times (1000 in our case) to further mitigate the shot noise. 
        \item For each star, the membership is computed as follows:
        \begin{equation}
            m_i = 1 - \frac{n_i^{sub}}{n_j^{total}}
        \end{equation}
        where $n_i^{sub}$ is the number of times that the i-th star has been subtracted, and $n_j^{total}$ is the total number of subtractions made in the j-th cell that contains the star.
    \end{enumerate}
    The process of statistical subtraction is combined with a differential reddening correction using the algorithm described in \cite{milone_2012}.
    We combine iteratively the differential reddening correction and the statistical subtraction through the following algorithm:
    \begin{enumerate}
        \item The differential reddening correction of the target CMD is performed using as reference stars only the ones with a membership probability higher than a certain threshold (we use $90\%$). For the first iteration all the stars are used, as their membership probability is 1.
        \item We perform the statistical decontamination of the CMD using the differential reddening corrected magnitudes;
        \item We iterate the two steps by updating the membership probabilities and the reddening correction.
    \end{enumerate}
    We consider that the process has reached convergence when the typical residual reddening corrections are of the order of the photometric error, which in this case happens after 3 iterations. 
    The result of our procedure is shown in Fig.~\ref{fig:membership_overview}. In the top panels, the reddening-corrected CMDs of NGC~6749 and of its parallel field are shown, respectively. From these, it is already evident that the overlap between the target main sequence (MS) and the field MS is quite strong, making this a challenging case for the statistical approach. The panel in the lower-left corner shows the same target CMD but colour-coded based on the membership probability, which in turn is displayed in the lower-right corner. The CMD of likely members appears very well defined, with likely contaminants populating the expected bluer and brighter sequence typically attributed to MW disc stars in the GC foreground or background.

    \subsection{Comparison between the two methods}\label{subsec:membership_comparison}
\begin{figure*}[tb]
            \centering
            \includegraphics[width=\textwidth]{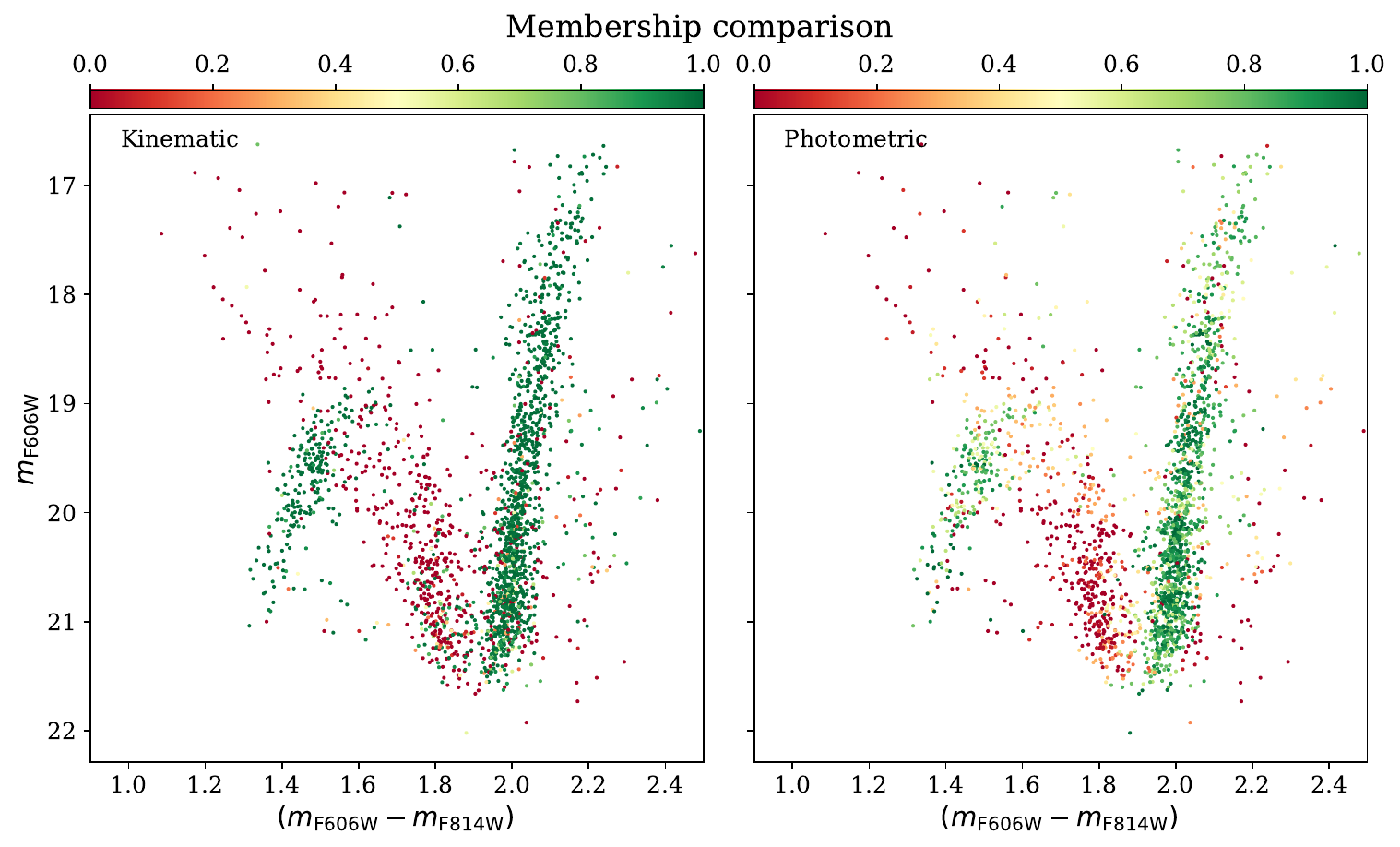}
            \caption{CMDs color coded for the membership probability: ({\it left}) membership calculated from the PMs, ({\it right}) membership obtained with the statistical approach.}
            \label{fig:membership_comparison}
    \end{figure*}

    Using stars that have both photometric and kinematic information, we make a direct comparison between the membership distributions obtained with the two methods described above. In Fig.~\ref{fig:membership_comparison} the target CMD is color-coded according to the membership probability obtained from the kinematic approach in the left-hand panel and the photometric one in the right-hand panel. The two results are remarkably similar, thus demonstrating that the photometric procedure has worked successfully.
    Since this method will be the primary one adopted by MGCS, on GCs with different properties, it is worth discussing some of the limitations that the statistical approach has, compared to the kinematic one. First, it is limited by the number of stars in the target CMD. In particular, few stars in the target CMD means bigger Voronoi cell size, thus lowering the resolution and the ability to sample small and isolated features. Second, the dilation of the Voronoi cells at the edges of features like the MS and the giant branches, merges some of the small cells mapping the feature itself with the large cells mapping the almost empty space next to it. As a result, these are more likely to be over-subtracted during the process. This mimics an {\it erosion} effect of the feature edges, where the membership probability is lower than in the inner parts. Finally, where no field signal is present in some cell (meaning that no stars are found in that cell in the field CMD), the membership is exactly 1, while when the field signal is dominating the target (i.e., when the number of field stars in that cell is greater than the target ones), the membership drops to 0. 
    However, the result shown in Fig.~\ref{fig:membership_overview} and Fig.~\ref{fig:membership_comparison} suggests that our statistical approach is a valid method to decontaminate the entire target CMD. Therefore, we advice any user to pick one method over the other depending on the needs of the specific science case of interest.
    
\section{Kinematic analysis}\label{sec:kinematic_analysis}
By using the unprecedentedly precise PMs that we measured, and the GMM described in Section~\ref{subsec:kinematic_membership}, we perform a detailed analysis of the internal kinematics of NGC~6749. Firstly, we validate our approach in two ways: $i$) with a mock catalog with known parameters (see Section~\ref{subsec:mock_validation}) and $ii$) with the \gaia DR3 PMs already presented in the literature. After validation, in Section~\ref{subsec:radial_analysis} we refine the mean PM of the GC and build the first PM-dispersion radial profile for NGC~6749.

    \subsection{Validation with a mock catalog and external checks}\label{subsec:mock_validation}
    We made a first validation of our GMM using a mock distribution of PMs, which qualitatively resembles the one shown in the VPD of NGC~6749, convolved with an error of $0.5\,\mathrm{mas}\,\,\mathrm{yr}^{-1}$. 
    Table~\ref{tab:mock_validation} summarizes the comparison between the posterior distributions of the GMM parameters, using the median value as reference and the 16th and 84th as uncertainties (i.e. the $1\sigma$ confidence range), and the parameters used to generate the mock distribution. The posterior distributions are always consistent within 1$\sigma$ with the true input values, except for the mean motion of the cluster and the field along Declination (the $\mu_{\delta}$ component), which are consistent with the input within 2$\sigma$. This is likely due to the fact that, given the shape of the VPD, the $\delta$ direction is the one most affected by the overlap between the cluster and the field component. 
    However, this test shows that our GMM can successfully recover the properties of a PM distribution where the field and cluster components have a strong overlap.
    \begin{table}
    \caption{Comparison between the posterior model distributions and the mock model.}
    \label{tab:mock_validation}
    \centering
    \begin{tabular}{c c c}
        \hline\hline
        Parameters & Posterior & Mock \\ 
        \hline
           $\mu_{\alpha}cos\delta^{GC}\,\,[\mathrm{mas}\,\,\mathrm{yr}^{-1}]$      &   $-3.03_{-0.03}^{+0.02}$ & $-3.00$\\
           $\mu_{\alpha}cos\delta^{Field}\,\,[\mathrm{mas}\,\,\mathrm{yr}^{-1}]$   &   $-1.68_{-0.15}^{+0.15}$ & $-1.75$\\
           $\mu_{\delta}^{GC}\,\,[\mathrm{mas}\,\,\mathrm{yr}^{-1}]$               &   $-5.45_{-0.02}^{+0.03}$ & $-5.50$\\
           $\mu_{\delta}^{Field}\,\,[\mathrm{mas}\,\,\mathrm{yr}^{-1}]$            &   $-4.51_{-0.14}^{+0.15}$ & $-4.25$\\
           $\sigma_{\alpha}^{GC}\,\,[\mathrm{mas}\,\,\mathrm{yr}^{-1}]$            &   $0.19_{-0.07}^{+0.05}$  & 0.21\\
           $\sigma_{\alpha}^{Field}\,\,[\mathrm{mas}\,\,\mathrm{yr}^{-1}]$         &   $2.50_{-0.11}^{+0.11}$  & 2.50\\
           $\sigma_{\delta}^{GC}\,\,[\mathrm{mas}\,\,\mathrm{yr}^{-1}]$            &   $0.20_{-0.07}^{+0.06}$  & 0.21\\
           $\sigma_{\delta}^{Field}\,\,[\mathrm{mas}\,\,\mathrm{yr}^{-1}]$         &   $2.58_{-0.12}^{+0.12}$  & 2.50\\
           $\rho_{\alpha\delta}^{GC}\,\,[\mathrm{-}]$                              &   $0.03_{-0.05}^{+0.05}$  & 0.0\\ 
           $\rho_{\alpha\delta}^{Field}\,\,[\mathrm{-}]$                           &   $0.25_{-0.05}^{+0.05}$  & 0.3\\ 
           $w_{GC}\,\,[\mathrm{-}]$                                                &   $0.65_{-0.02}^{+0.02}$  & 0.67\\ 
           $w_{Field}\,\,[\mathrm{-}]$                                             &   $0.35_{-0.02}^{+0.02}$  & 0.33\\ 
        \hline
    \end{tabular}
    \end{table}


    After the validation with the mock PMs, we tested our approach on the sample of member stars of NGC~6749 used to obtain the value of velocity dispersion quoted in \cite{vasiliev_2021}. We remark that in this test ({\it VB21} hereafter), given that the sample is composed only by member stars, we used a single-component GMM.
    
\begin{figure}[htb]
        \centering
        \includegraphics[width=\columnwidth]{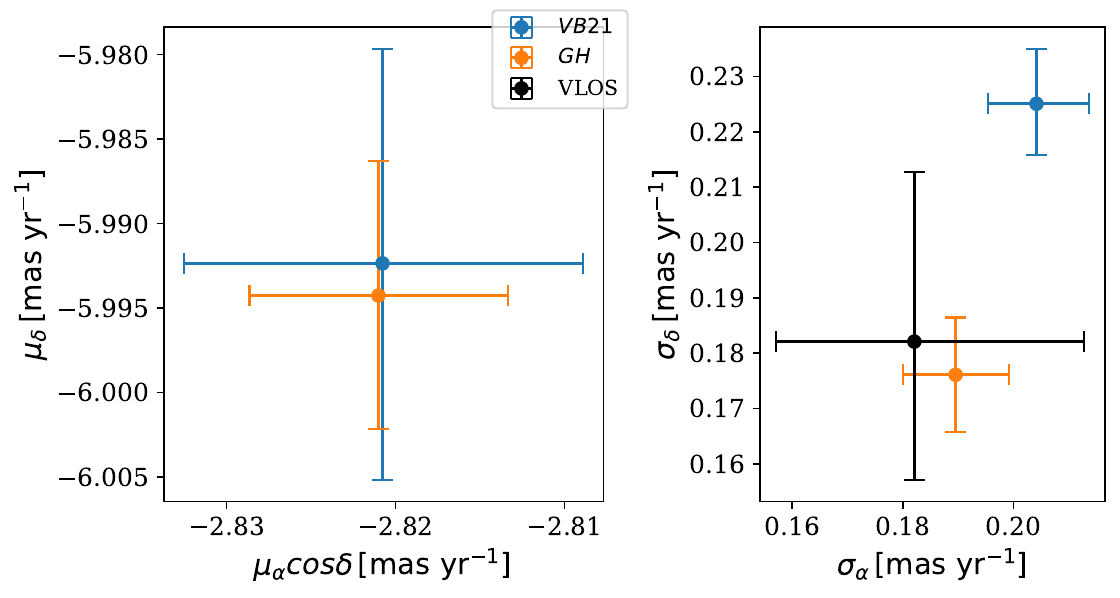}
        \caption{Comparison of the mean velocity ($left$) and velocity dispersion ($right$) components for \gh (orange), and \gaia{DR3}{\it-VB21}(blue). The errorbar are the 16th and 84th of the parameter's posterior distribution. The up-to-date line-of-sight velocity dispersion (black) is shown as reference.}
        \label{fig:parameters_compariosn}
    \end{figure}

    Fig.~\ref{fig:parameters_compariosn} shows the comparison of the cluster's mean motion components $\mu_{\alpha,\delta}$ and the PM dispersions $\sigma_{\alpha,\delta}$ among the two PMs distributions. As can be seen from this figure, the $\mu_{\alpha,\delta}$ of the cluster component is well consistent between the two cases. Our measurement provides a mean PM of $\mu_{\alpha,\delta}=(-2.82, -5.99)\, \mathrm{mas}\,\,\mathrm{yr}^{-1}$,  with a smaller uncertainty ($0.008\,\mathrm{mas}\,\,\mathrm{yr}^{-1}$) with respect to {\it VB21} ($0.012\,\mathrm{mas}\,\,\mathrm{yr}^{-1}$). On the other hand, we find a larger discrepancy (at the 3$\sigma$ level) in $\sigma_{\delta}$, with our \gh PMs providing $\sigma_{\delta}=0.18\pm0.01\,\mathrm{mas}\,\,\mathrm{yr}^{-1}$, while for the {\it VB21} sample we determine $\sigma_{\delta}=0.23\pm0.01\,\mathrm{mas}\,\,\mathrm{yr}^{-1}$. 

    To understand which of the two estimates is more realistic, we compared with the up to date line-of-sight velocity dispersion values available for NGC~6749\footnote{Baumgardt H., private communication.}, i.e., $\sigma_{VLOS}=6.55_{-0.90}^{+1.10}$~km/s as measured with a combination of Gaia DR3, Apogee DR19 and published literature RVs.
    Under the assumption of isotropy, our \gh velocity dispersion estimates of $\sigma_\alpha=6.83$ km/s and $\sigma_{\delta}=6.34$ km/s (assuming a distance of 7.59 kpc, \citealt{baumgardt21}) agree better than the ones we get from {\it VB21} of $\sigma_\alpha=7.35$ km/s and $\sigma_{\delta}=8.11$ km/s. This is reasonable evidence that our measurements led to an improved estimate of the cluster velocity dispersion. Likely, our more precise \gh PMs allow for a better deconvolution of cluster and field components in the GMM, thus enabling a better selection of the actual stellar members.
    In passing, we also provide an estimate for the kinematic distance (d$_{kin}$) of NGC~6749. Under the usual assumption of isotropy, the tangential velocity dispersion that we measure matches that along the line-of-sight for d$_{kin}=6.92\pm0.36$ kpc, which is consistent within 1.2$\sigma$ with the distance quoted by \cite{baumgardt21} of d$=7.59\pm0.22$ kpc.
    
    \subsection{Kinematic analysis of NGC~6749}\label{subsec:radial_analysis}
\begin{figure}[htb]
        \centering
        \includegraphics[width=\columnwidth]{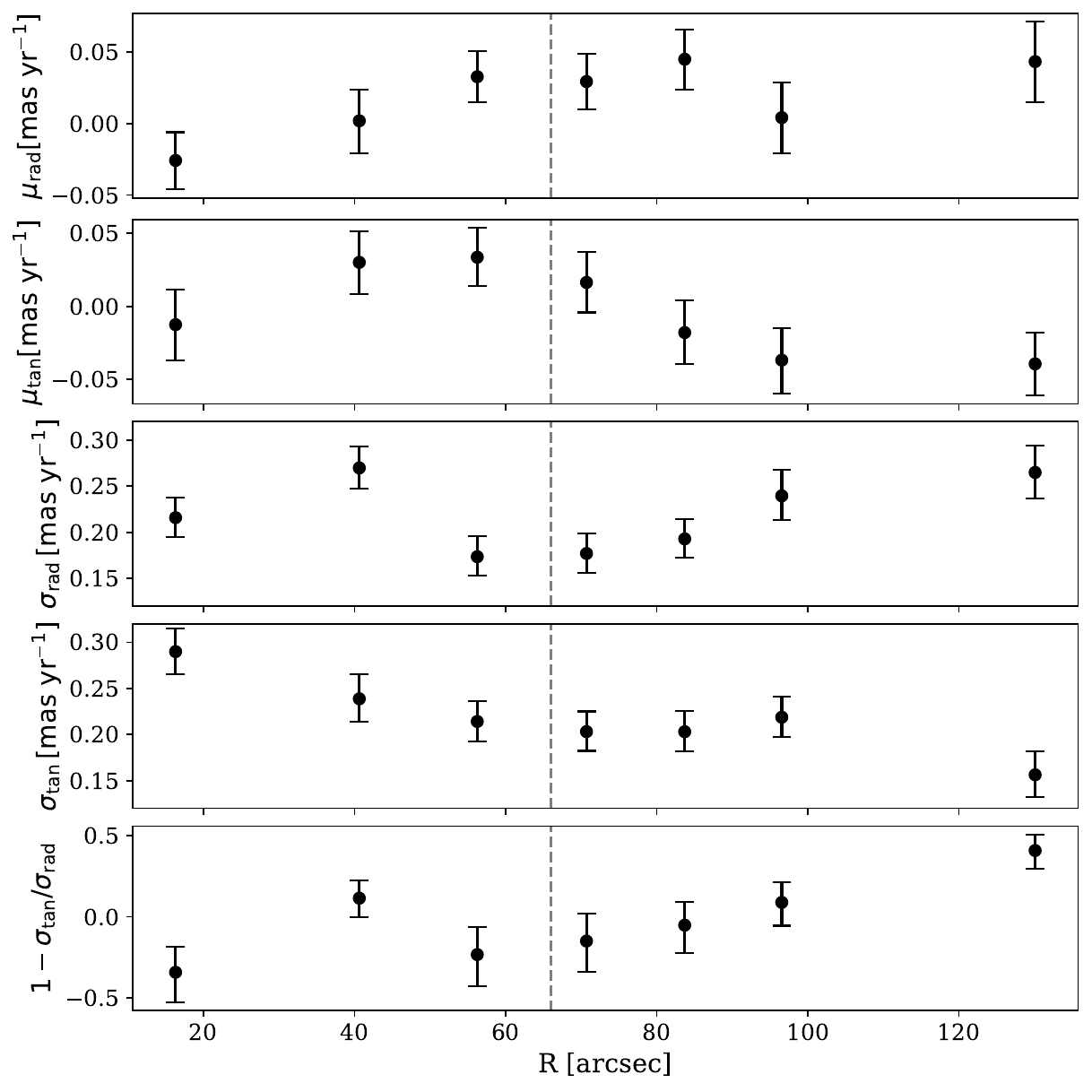}
        \caption{Radial profiles of the mean velocity ($\mu_{\mathrm{rad,tan}}$), velocity dispersion ($\sigma_{\mathrm{rad,tan}}$) and anisotropy (calculated as $1-\frac{\sigma_{tan}}{\sigma_{rad}}$)for NGC~6749. The half-light radius (gray dashed line) is shown as reference.}
        \label{fig:rad_tan_profiles}
    \end{figure}

    Thanks to the large number of stars with measured PMs available, we can now investigate the internal kinematics of NGC~6749. We divided our field in annular bins of variable area, each containing at least 300 stars.
    We also decomposed our PMs from Equatorial coordinates to their radial and tangential components, so to study features like kinematic anisotropies and overall expansion/rotation in the plane of the sky.  
    As we already point out in Section~\ref{subsec:data_red_astrometry}, our \gh PMs are more precise than \textit{Gaia}'s at its faint end ($G \gtrsim 18.5$) , while for brighter stars ($G \lesssim 18.5$) \gaia is still the best option. For this reason, we decided to construct a {\it golden sample} of stars by selecting the PMs that have the lowest uncertainty, either from \gh or \gaia. 
    For each bin, we modeled the VPD using the GMM described in Section~\ref{subsec:kinematic_membership}, by fixing the parameters of the field component to the median value of its posterior distribution, obtained using the {\it golden sample}. 
    Figure~\ref{fig:rad_tan_profiles} reports the radial trends of the cluster's mean PM and PM dispersions along the radial and tangential directions $\sigma_{rad,tan}$.
    We also measure the ratio between the radial and tangential components of the velocity dispersion, thus obtaining the anisotropy profile of NGC~6749. 
    The trend of the mean GC motion as a function of radius can be indicative of kinematic features such as rotation or expansion/contraction. In our case, the best-fit linear trends gives an angular coefficient of ($5.0\pm2.2\cdot10^{-4}$) $\mathrm{mas}\,\,\mathrm{yr}^{-1}\,\,\mathrm{arcsec}^{-1}$  in the radial direction and ($5.0\pm2.9\cdot10^{-4}$) $\mathrm{mas}\,\,\mathrm{yr}^{-1}\,\,\mathrm{arcsec}^{-1}$ in the tangential one. These numbers might indicate mild systematic motions, particularly in the radial direction, where the trend is inconsistent with 0 at 2.2$\sigma$. We have also checked that, given the cluster distance, v$_{LOS}$\footnote{Quoted in \href{https://people.smp.uq.edu.au/HolgerBaumgardt/globular/orbits.html}{people.smp.uq.edu.au/HolgerBaumgardt/globular/orbits}} and FoV covered by our data, any correction for perspective effects is of the order of $10^{-3}\,\,\mathrm{mas}\,\mathrm{yr}^{-1}$, and is therefore negligible compared to the magnitude of the observed trends. However, given the relatively small field of view covered by our PMs, we prefer to avoid strong claims and leave this as a possible evidence to be further investigated. 

    As for the PM dispersion profile, we detect a rather flat behavior in the radial direction, and a decreasing trend along the tangential component. The resulting anisotropy profile shows a slightly increasing trend, with isotropy holding out to about 100 arcsec, and a mild radial anisotropy at distance larger than $\sim120$". Considering the values from \citet[2010 edition; 37" and 66" for core and half-light radii, respectively]{harris96}, our anisotropy profile is qualitatively in agreement with that of GCs with an intermediate dynamical age shown by \citet{libralato_2022}, but its velocity-dispersion radial profile is steeper than that of GCs with the same concentration index ($c=0.79$). A more detailed analysis of the structural properties of this GC, thanks to the exquisite new \hst data available, is necessary to further investigate the system. 

    \subsection{Comparison with simulations}\label{subsec:nbody}

\begin{figure*}[htb]
        \centering
        \includegraphics[width=\textwidth]{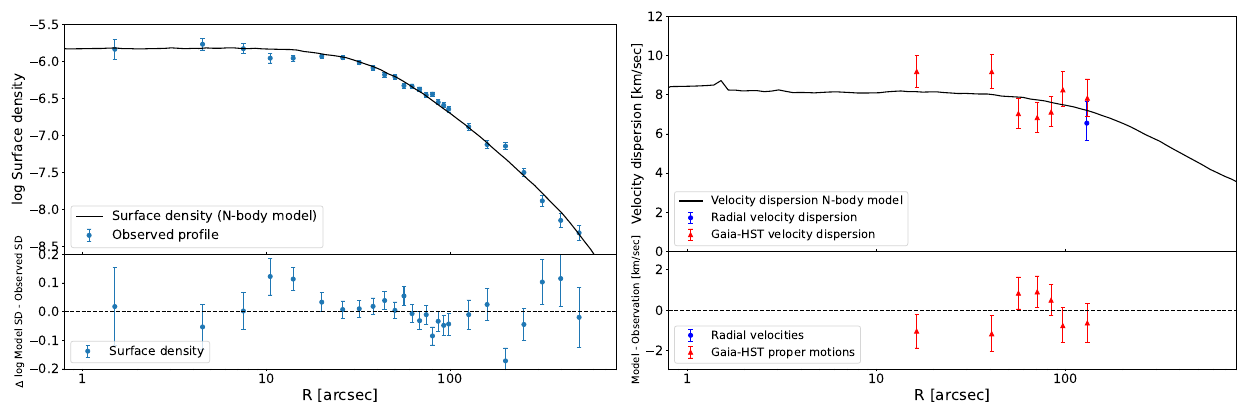}
        \caption{Comparison of the surface density profile ($left$) and the velocity dispersion profile ($right$) of NGC~6749 with the best-fitting $N$-body model from the grid of models from \citet{baumgardt18}. We obtain good agreement between the model and the observations.}
        \label{fig:nbody}
    \end{figure*}

In order to derive the physical parameters of NGC~6749 from the new velocity dispersion profile, we searched the grid of $N$-body simulations that produces the best match of the surface density and velocity dispersion profile of NGC~6749 from \citet{baumgardt18}. In order to compare with the $N$-body simulations, we summed the observed radial and tangential velocity dispersions at each radius, which seems justified since the velocity dispersion is close to isotropic over most of the radial range that we have measured it. We scaled all $N$-body simulations in radius to match the observed half-light radius of NGC~6749 and then interpolated in the grid of simulations to find the best-fitting $N$-body model. We then derived the physical parameters of NGC~6749 from the $N$-body simulations as described in \citet{baumgardt18}. Fig.~\ref{fig:nbody} shows the result of our fitting process. It can be seen that we obtain a good fit to both the velocity dispersion and surface density of NGC~6749 and that our proper motion velocity dispersion profile is in good agreement with the radial velocity dispersion profile for an assumed cluster distance of 7.6 kpc. We find a total cluster mass of $M_C=(5.53 \pm 0.28)\cdot 10^5$ M$_\odot$, a projected central dispersion of $\sigma_0=8.0\pm0.64$ km~s$^{-1}$, an $M/L$ ratio of $M/L=3.65 \pm 0.27 M_\odot/L_\odot$ and a half-mass radius of $r_h=7.1$ pc, putting NGC~6749 among the 10\% most massive globular clusters in the Milky Way. 

\color{black}

\section{Conclusions}\label{sec:conclusions}
As a part of the MGCS treasury program \citep{massari_2025mgcs}, we present in this paper new \hst observations of NGC~6749 that we used to develop new, publicly available python-based tools for the photometric-based membership determination and kinematic analysis of GCs (and GC-like objects).

By combining our MGCS \hst observations \citep{massari_2025mgcs} with \gaia DR3, using the software \texttt{GAIAHUB}, we determined unprecedentedly precise PMs for this GC. In particular, we are able to decrease the PM error by up to one order of magnitude the faint end ($G > 18.5$), and for the first time we determine PMs for stars that only had {\it Gaia} positions measured so far.

We analysed the resulting VPD of NGC~6749 using GMM. This method comprises of an inference-based approach that uses MCMC simulations to find the posterior distribution of the model parameters. Our model has the flexibility of managing an arbitrary number of gaussian components, as well as the ability to fix and/or link together any parameters to each other. A natural product of this procedure is the probability of each star to belong to NGC~6749. We compared this kinematic membership with that stemming out from a photometric approach that combines a statistical CMD subtraction, featured with an adaptive Voronoi-based grid generation, with correction for differential reddening. Such an an iterative approach improves the membership probability and the differential reddening correction at each iteration. The comparison between the two methods is remarkably good, with the photometric algorithm ensuring an appropriate CMD decontamination even for stars too faint to have PM information.

Our kinematic analysis is first validated with a mock catalogue, and then with existing kinematic measurements, thus verifying the efficacy of our method in separating cluster and field components even in a relatively highly contaminated case. The mean motion of NGC~6749 of $\mu_{\alpha,\delta}=(-2.82, -5.99)\pm0.008\, \mathrm{mas}\,\,\mathrm{yr}^{-1}$, is determined with a precision better than {\it Gaia}. Given the large number of stars with a precise PM measurement, we could investigate for the first time the kinematic profiles of NGC~6749. By using a golden sample of PMs, that combines the best measurements from ours and {\it Gaia} catalogues, we built the profiles of $\mu_{rad, tan}$, $\sigma_{rad, tan}$ and anisotropy ($\sigma_{tan}/\sigma_{rad}$). We found an overall agreement with the general kinematic picture of GCs drawn by \hst\citep{libralato_hubble_2022}. Finally, the comparison of our observed kinematic profile with N-body simulations revealed that NGC~6749 is among the the 10\% most massive GC in the Galaxy, with a total mass of $M=(5.53 \pm 0.28)\cdot 10^5$ M$_\odot$.

\section*{Data availability}
All the tools presented here are available at the GitHub repository \url{https://github.com/LR-inaf/MGCS_pytools.git} or at the PyPi repository \url{https://pypi.org/project/MGCS-pytools/}.

\appendix
\section{Systematic correction}\label{appendix:pms_calibration}
\begin{figure}[htb]    
    \centering
    \begin{subfigure}[b]{0.5\textwidth}
        \centering
        \includegraphics[width=\textwidth]{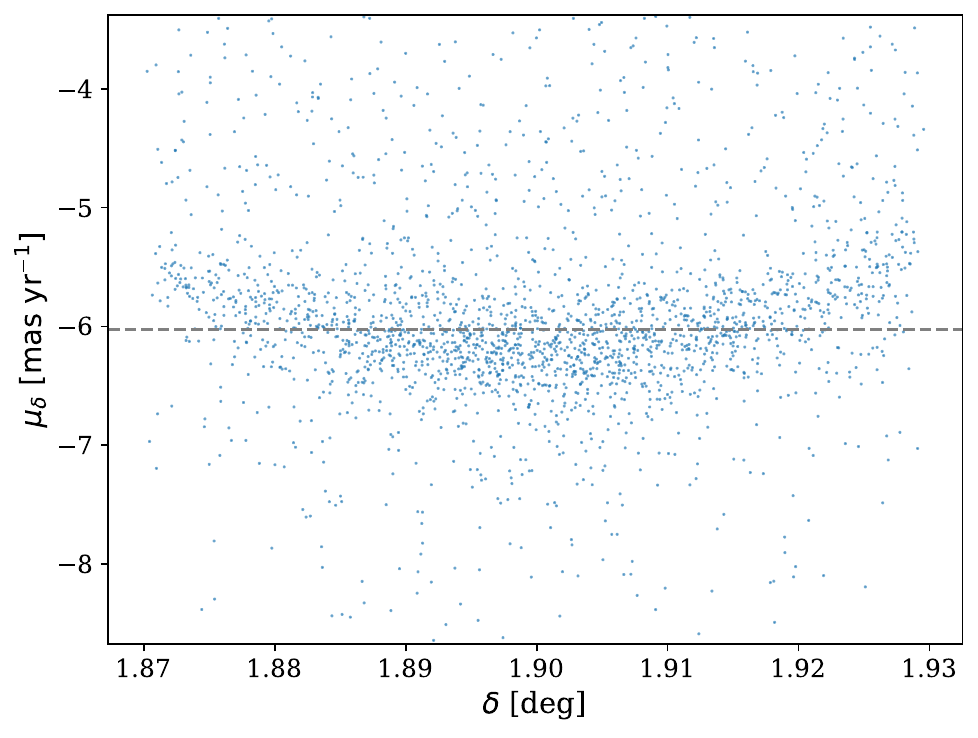}
        \caption{}
        \label{fig:pms_systematic_uncorr}
    \end{subfigure}
    \hfill
    \begin{subfigure}[b]{0.5\textwidth}
        \centering
        \includegraphics[width=\textwidth]{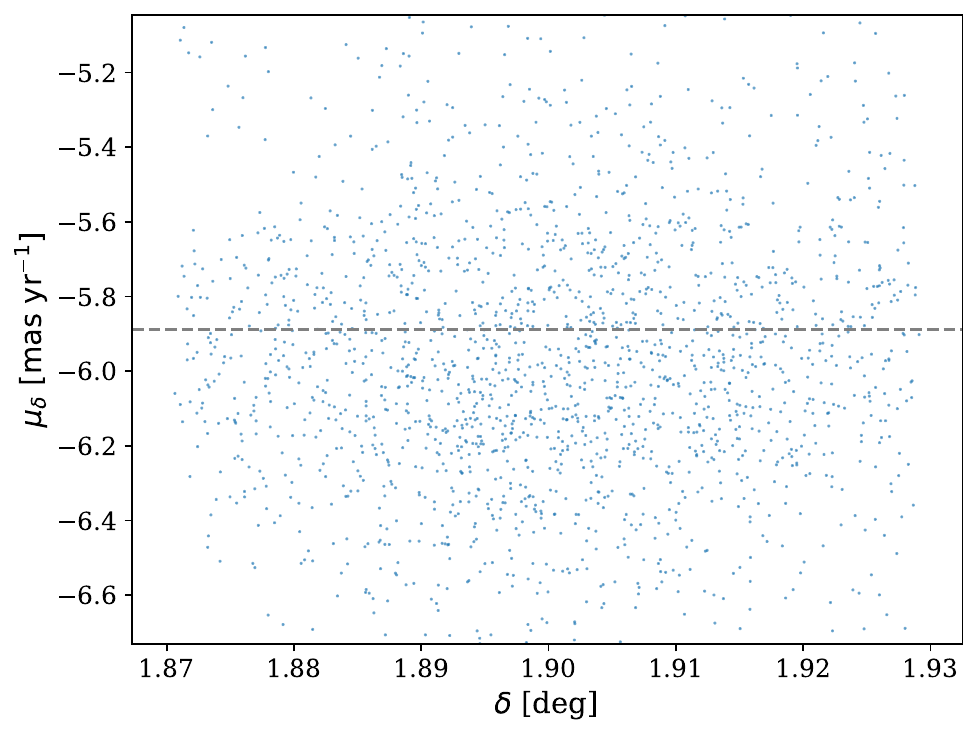}
        \caption{}
        \label{fig:pms_systematic_corr}
    \end{subfigure}
    \caption{\gh PMs $\delta$ component against the sky $\delta$ coordinate. (a) The uncorrected PMs and (b) after the correction. The gray dashed line shows the median value of the distribution, used as visual reference.}
    \label{fig:pms_systematic}
\end{figure}

We extensively checked for systematics introduced by \texttt{GAIAHUB} and we found a non-linear correlation between $\mu_{\delta}$ and the position of stars in the field of view, in particular with the $\delta$ coordinate (see Fig. \ref{fig:pms_systematic_uncorr}). Firstly, we cross-checked this relation with the pure \gaia DR3 PM and we did not found any significant relation, reinforcing the idea of having introduced a systematic using \texttt{GAIAHUB}. We then choose to use \gaia as reference to compensate for this dependence. We corrected the \gh PM of each star by subtracting the difference between itself and the median value of the \gaia PMs of the 25 neighbors stars. With this approach we were able to remove the systematic, obtaining an uncorrelated $\mu_{\delta}$ against $\delta$ (see Fig. \ref{fig:pms_systematic_corr}).

\begin{acknowledgements}

DM, SC, EP and AM acknowledge financial support from PRIN-MIUR-22: CHRONOS: adjusting the clock(s) to unveil the CHRONO-chemo-dynamical Structure of the Galaxy” (PI: S. Cassisi). A.M. and M.B. acknowledge support from the project "LEGO – Reconstructing the building blocks of the Galaxy by chemical tagging" (P.I. A. Mucciarelli), granted by the Italian MUR through contract PRIN 2022LLP8TK\_001.
M.M. acknowledges support from the Agencia Estatal de Investigaci\'on del Ministerio de Ciencia e Innovaci\'on (MCIN/AEI) under the grant "RR Lyrae stars, a lighthouse to distant galaxies and early galaxy evolution" and the European Regional Development Fun (ERDF) with reference PID2021-127042OB-I00.
Co-funded by the European Union (ERC-2022-AdG, "StarDance: the non-canonical evolution of stars in clusters", Grant Agreement 101093572, PI: E. Pancino). Views and opinions expressed are however those of the author(s) only and do not necessarily reflect those of the European Union or the European Research Council. Neither the European Union nor the granting authority can be held responsible for them.
This paper is supported by the Italian Research Center on High Performance Computing Big Data and Quantum Computing (ICSC), project funded by European Union - NextGenerationEU - and National Recovery and Resilience Plan (NRRP) - Mission 4 Component 2 within the activities of Spoke 3 (Astrophysics and Cosmos Observations).
This work is part of the project Cosmic-Lab at the Physics and Astronomy Department “A. Righi” of the Bologna University (\url{http:// www.cosmic-lab.eu/ Cosmic-Lab/Home.html}).

Based on observations with the NASA/ESA HST, obtained
at the Space Telescope Science Institute, which is operated by
AURA, Inc., under NASA contract NAS 5-26555. Support for Program number GO-17435 was provided through grants from STScI under NASA contract NAS5-26555. This
research made use of emcee \citep{Foreman-Mackey_2013}.
This work has made use of data from the European Space Agency (ESA) mission
{\it Gaia}\ (\url{https://www.cosmos.esa.int/gaia}), processed by the {\it Gaia}\
Data Processing and Analysis Consortium (DPAC,
\url{https://www.cosmos.esa.int/web/gaia/dpac/consortium}). Funding for the DPAC
has been provided by national institutions, in particular the institutions
participating in the {\it Gaia}\ Multilateral Agreement.

ED acknowledges financial support from the INAF Data Analysis Research Grant (PI E. Dalessandro) of the “Bando Astrofisica Fondamentale 2024”.
\end{acknowledgements}

%
%

\bibliographystyle{aa} 
\bibliography{aa} 

\end{document}